\documentclass[a4paper]{jpconf}
\usepackage{graphicx}
\begin{document}
\title{Basics of inflationary cosmology}

\author{George Lazarides}

\address{Physics Division, School of Technology,
Aristotle University of Thessaloniki, Thessaloniki 54124, Greece}

\ead{lazaride@eng.auth.gr}

\begin{abstract}
The early stages of the universe evolution are discussed
according to the hot big bang model and the grand unified
theories. The shortcomings of big bang are summarized and their
resolution by inflationary cosmology is sketched. Cosmological
inflation, the subsequent oscillation and decay of the inflaton
field, and the resulting reheating of the universe are studied
in some detail. The density perturbations produced by inflation
and the temperature fluctuations of the cosmic microwave
background radiation are introduced. The hybrid inflationary model
is described. Two natural variants of the supersymmetric version
of this model which avoid the disaster encountered in its standard
realization from the overproduction of magnetic monopoles are
presented.
\end{abstract}

\section{Introduction}
\label{sec:introduction}

\par
The discovery of the cosmic microwave background radiation (CMBR)
together with the observed Hubble expansion of the universe had
established hot big bang as a viable model of the universe (for
a textbook treatment of this model, see e.g. Ref.~\cite{wkt}).
The success of nucleosynthesis (see e.g. Ref.~\cite{bbn}) in
reproducing the observed abundance of light elements in the
universe and the proof of the black body character of the CMBR
then imposed hot big bang as the standard cosmological model.
This model combined with the grand unified theories (GUTs)
\cite{ggps} of strong, weak, and electromagnetic interactions
provides the scientific framework for discussing the early
stages of the universe evolution.

\par
The standard big bang (SBB) cosmological model, despite its
great successes, had some long-standing shortcomings. One of
them is the so-called horizon problem. The CMBR received now
has been emitted from regions of the universe which, according
to this model, had never communicated before sending this
radiation to us. The question then arises how come the
temperature of the
black body radiation from these regions is so finely tuned as
the measurements of the cosmic background explorer (COBE)
\cite{cobe} and the Wilkinson microwave anisotropy probe
(WMAP) \cite{wmap1,wmap3} satellites show. Another important
issue is the so-called flatness problem. It is a fact
\cite{wmap3} that the present universe appears to be almost
flat. This means that, in its early stages, the universe must
have been flat with a great accuracy, which requires an extreme
fine tuning of its initial conditions. Also, combined with GUTs
which predict the existence of superheavy magnetic monopoles
\cite{monopole}, the SBB model leads \cite{preskill} to a
catastrophe caused by the overproduction of these monopoles.
Finally, the model has no explanation for the small density
perturbations which are required for explaining the structure
formation in the universe (for a pedagogical discussion, see
e.g. Ref.~\cite{structure}) and the generation of the observed
\cite{cobe,wmap1,wmap3} temperature fluctuations in the CMBR.

\par
It is clear that cosmological inflation \cite{guth} offers an
elegant solution to all these problems of the SBB model (for a
textbook introduction or previous reviews on inflation, see e.g.
Refs.~\cite{book,inflation}). The idea is that, in the early
universe, a real scalar field (the inflaton) was displaced
from its vacuum value. If the potential energy density of this
field happens to be quite flat, the roll-over of the field towards
the vacuum can be very slow for a period of time. During this
period, the energy density is dominated by the almost constant
potential energy density of the inflaton. As a consequence, the
universe undergoes a period of quasi-exponential expansion, which
can readily solve the horizon and flatness problems by stretching
the distance over which causal contact is established and reducing
any pre-existing curvature in the universe. It can also adequately
dilute the GUT magnetic monopoles. Moreover, it provides us with the
primordial density perturbations which are needed for explaining
the large scale structure in the universe \cite{structure} as well
as the temperature fluctuations observed in the CMBR. Inflation
can be easily incorporated in GUTs. It can occur during the GUT
phase transition at which the GUT gauge symmetry breaks by the
vacuum expectation value (VEV) of a Higgs field, which also plays
the role of the inflaton.

\par
After the termination of inflation, the inflaton field starts
performing oscillations about the vacuum. These oscillations are
damped because of the dilution of the field energy density by the
cosmological expansion and the decay of the inflaton into light
particles. The resulting radiation energy density eventually
dominates over the field energy density and the universe
returns to a normal big bang type evolution. The temperature
at which this occurs is historically called `reheat'
temperature although there is neither supercooling nor
reheating of the universe \cite{reheat} (see also
Ref.~\cite{dilution}).

\par
The early realizations of inflation share the following important
disadvantage. They require tiny parameters in order to reproduce
the COBE or WMAP measurements on the CMBR. In order to solve this
naturalness problem, hybrid inflation has been introduced
\cite{hybrid}. The basic idea was to use two real scalar fields
instead of one that was customarily used. One field may be a gauge
non-singlet and provides the `vacuum' energy density which drives
inflation, while the other is the slowly varying field during
inflation. This splitting of roles between two fields allows us to
reproduce the temperature fluctuations of the CMBR with natural
(not too small) values of the relevant parameters. Hybrid
inflation, although it was initially introduced in the context of
non-supersymmetric GUTs, can be naturally incorporated
\cite{lyth,dss} in supersymmetric (SUSY) GUTs.

\par
It is unfortunate that the magnetic monopole problem reappears in
hybrid inflation. The termination of inflation, in this case, is
abrupt and is followed by a `waterfall' regime during which the
system falls towards the vacuum manifold and performs damped
oscillations about it. If the vacuum manifold happens to be
homotopically non-trivial, topological defects will be copiously
formed \cite{smooth} by the Kibble mechanism \cite{kibble} since
the system can end up at any point of this manifold with equal
probability. Therefore, a cosmological disaster is encountered in
the hybrid inflationary models which are based on a gauge symmetry
breaking predicting magnetic monopoles.

\par
One way \cite{smooth,jean,shi,talks} to solve the magnetic monopole
problem of SUSY hybrid inflation is to include into the standard
superpotential for hybrid inflation the leading non-renormalizable
term. This term cannot be excluded by any symmetries and, if its
dimensionless coefficient is of order unity, can be comparable
with the trilinear coupling of the standard superpotential (whose
coefficient is typically $\sim 10^{-3}$). Actually, we have two
options. We can either keep \cite{jean} both these terms or remove
\cite{smooth,shi} the trilinear term by imposing a discrete
symmetry and keep only the leading non-renormalizable term. The
pictures emerging in the two cases are quite different. However,
they share an important common feature. The GUT gauge group is
spontaneously broken already during inflation and, thus, no
topological defects can form at the end of inflation. So, the
magnetic monopole problem is solved.

\section{The Big Bang Model}
\label{sec:bigbang}

\par
We will start with an introduction to the salient features of the
SBB model \cite{wkt} and a summary of the history of the early
universe in accordance to GUTs.

\subsection{Hubble Expansion}
\label{subsec:hubble}

\par
At cosmic times $t\stackrel{_{>}}{_{\sim }} t_{\rm P}\equiv
M_{\rm P}^{-1}\sim 10^{-44}~{\rm{sec}}$ ($M_{\rm P}=1.22\times
10^{19}~{\rm{GeV}}$ is the Planck scale) after the big bang, the
quantum fluctuations of gravity are suppressed and classical
general relativity yields an adequate description of gravity.
Strong, weak, and electromagnetic interactions, however, require
quantum field theoretic treatment.

\par
We assume that the universe is homogeneous and isotropic. The
strongest evidence for this {\it cosmological principle} is the
observed \cite{cobe,wmap1,wmap3} isotropy of the CMBR. The
space-time metric then takes the Robertson-Walker form
\begin{equation}
ds^{2}=-dt^{2}+ a^{2}(t)\left[\frac{dr^{2}}
{1-kr^2}+r^{2}(d\vartheta^{2}+\sin^{2}\vartheta~ d\varphi^{2})
\right],
\label{eq:rw}
\end{equation}
where $r$, $\varphi$, and $\vartheta$ are `comoving' polar
coordinates, which remain fixed for objects that just follow the
general cosmological expansion. The parameter $k$ is the scalar
curvature of the 3-space and $k=0$, $>0$, or $<0$ corresponds to
flat, closed, or open universe. The dimensionless parameter $a(t)$
is the scale factor of the universe. We take
$a_{0}\equiv a(t_{0})=1$, where $t_{0}$ is the present value of
the cosmic time.

\par
The instantaneous radial physical distance is given by
\begin{equation}
R=a(t)\int_{0}^{r}\frac{dr}{(1-kr^{2})^\frac{1}{2}}.
\label{eq:dist}
\end{equation}
For flat universe ($k=0$), $\bar{R}=a(t)\bar{r}$ ($\bar{r}$ is a
comoving and  $\bar{R}$ a physical radial vector in 3-space) and
the velocity of an object is
\begin{equation}
\bar{V}=\frac{d\bar{R}}{dt}=\frac{\dot{a}}{a}\bar{R}
+a\frac{d\bar{r}}{dt},
\label{eq:velocity}
\end{equation}
where overdots denote derivation with respect to $t$. The second
term in the right hand side (RHS) of this equation is the
peculiar velocity, $\bar{v}= a(t) \dot{\bar{r}}$, of the object
i.e. its velocity with respect to the comoving coordinate
system. For $\bar{v}=0$, Eq.~(\ref{eq:velocity}) becomes
\begin{equation}
\bar{V}=\frac{\dot{a}}{a}\bar{R}\equiv H(t)\bar{R},
\label{eq:hubblelaw}
\end{equation}
where $H(t)\equiv\dot{a}(t)/a(t)$ is the Hubble parameter. This is
the well-known Hubble law asserting that all objects run away from
each other with velocities proportional to their distances and is
the first success of SBB cosmology.

\subsection{Friedmann Equation}
\label{subsec:friedmann}

\par
In a homogeneous and isotropic universe, the energy-momentum
tensor takes the form $(T_{\mu}^{~\nu})={\rm{diag}}(-\rho, p, p,
p)$, where $\rho$ is the energy density, $p$ the pressure, and
the indices $\mu,~\nu=0,1,2,3$ correspond to the space-time
coordinates. Energy-momentum conservation then yields the
continuity equation
\begin{equation}
\frac{d\rho}{dt}=-3H(t)(\rho+p),
\label{eq:continuity}
\end{equation}
where the first term in the RHS describes the dilution of the
energy due to the Hubble expansion and the second term the work
done by pressure.

\par
For a universe described by the metric in Eq.~(\ref{eq:rw}),
Einstein's equations
\begin{equation}
R_{\mu}^{~\nu}-\frac{1}{2}~\delta_{\mu}^{~\nu}R= 8\pi
G~T_{\mu}^{~\nu},
\label{eq:einstein}
\end{equation}
where $R_{\mu}^{~\nu}$ and $R$ are the Ricci tensor and scalar
curvature respectively, $\delta_{\mu}^{~\nu}$ is the
Kronecker delta, and $G\equiv M_{\rm P}^{-2}$ is the
Newton's constant, lead to the Friedmann equation
\begin{equation}
H^{2}\equiv \left(\frac{\dot{a}(t)}{a(t)}\right)^{2} =\frac{8\pi
G}{3}\rho-\frac{k}{a^{2}}.
\label{eq:friedmann}
\end{equation}

\par
Averaging the pressure $p$, we write $\rho+p=(1+w)\rho\equiv
\gamma\rho$ and Eq.~(\ref{eq:continuity}) gives $\rho\propto
a^{-3\gamma}$. For a universe which is dominated by pressureless
matter, $\gamma=1$ and, thus, $\rho\propto a^{-3}$. This is
interpreted as mere dilution of a fixed number of particles in a
comoving volume due to the Hubble expansion of the universe. For
a radiation dominated universe, $p=\rho/3$ and, thus,
$\gamma=4/3$, which gives $\rho\propto a^{-4}$. The extra factor
of $a(t)$ is due to the red-shifting of all wave lengths by the
cosmological expansion. Substituting $\rho\propto a^{-3\gamma}$
in Eq.~(\ref{eq:friedmann}) with $k=0$, we get $a(t)\propto
t^{2/3\gamma}$ which, for $a(t_{0})=1$, gives
\begin{equation}
a(t)=\left(\frac{t}{t_0}\right)^\frac{2}{3\gamma}.
\label{eq:expan}
\end{equation}
For matter or radiation, we obtain $a(t)=(t/t_{0})^{2/3}$ or
$a(t)=(t/t_{0})^{1/2}$ respectively. So, we see that a matter
dominated universe expands faster than a radiation dominated
one.

\par
The early universe is radiation dominated and its energy density
is
\begin{equation}
\rho=\frac{\pi^{2}}{30}\left(N_{b}+\frac{7}{8}
N_{f}\right)T^{4}\equiv c~T^{4},
\label{eq:boltzman}
\end{equation}
where $T$ is the cosmic temperature and $N_{b(f)}$ the number of
massless bosonic (fermionic) degrees of freedom. The quantity
$g_{*}=N_{b}+(7/8)N_{f}$ is called effective number of massless
degrees of freedom. The entropy density is
\begin{equation}
s=\frac{2\pi^{2}}{45}~g_{*}~T^{3}.
\label{eq:entropy}
\end{equation}
Assuming adiabatic universe evolution i.e. constant entropy in a
comoving volume ($sa^{3}={\rm{constant}}$), we obtain
$aT={\rm{constant}}$. The temperature-time relation during
radiation dominance is then derived from Eq.~(\ref{eq:friedmann})
(with $k=0$):
\begin{equation}
T^{2}=\frac{M_{\rm P}}{2(8\pi c/3)^\frac{1}{2}t}.
\label{eq:temptime}
\end{equation}
Classically, the expansion starts at $t=0$ with $T=\infty$ and
$a=0$. This initial singularity is, however, not physical since
general relativity fails for $t\stackrel{_{<}}{_{\sim }}
t_{\rm P}$ (the Planck time). The only meaningful statement is
that the universe, after a yet unknown initial stage, emerges at
$t\sim t_{\rm P}$ with $T\sim M_{\rm P}$.

\subsection{Important Cosmological Parameters}
\label{subsec:parameter}

\par
The most important parameters which describe the expanding
universe are the following:
\begin{list}
\setlength{\rightmargin=0cm}{\leftmargin=0cm}
\item[{\bf i.}] The present value of the Hubble
parameter (known as Hubble constant) $H_{0} \equiv
H(t_{0})=100~h~\rm{km}~\rm{sec}^{-1}~ \rm{Mpc}^{-1}$ ($h=
0.72\pm 0.07$ from the Hubble space telescope \cite{h}).
\vspace{.25cm}
\item[{\bf ii.}] The fraction $\Omega=\rho/\rho_{c}$, where
$\rho_{c}$ is the critical energy density corresponding
to a flat universe. From Eq.~(\ref{eq:friedmann}),
$\rho_{c}=3H^{2}/8\pi G$ and $\Omega=1+k/a^{2}H^{2}$. $\Omega=1$,
$>1$, or $<1$ corresponds to flat, closed or open universe.
Assuming inflation (see below), the present value of $\Omega$ must
be $\Omega_{0}=1$ in accord with the DASI observations which yield
\cite{dasi} $\Omega_{0}=1.04\pm 0.08$. The low deuterium
abundance measurements \cite{deuterium} in view of nucleosynthesis
(see e.g. Ref.~\cite{bbn}) give $\Omega_{B}h^2=
0.020\pm 0.001$, where $\Omega_{B}$ is the baryonic contribution
to $\Omega_0$. This result implies that $\Omega_{B}\approx
0.039$. The total contribution $\Omega_{\rm m}$ of matter to
$\Omega_0$ can then be determined from the measurements
\cite{cluster} of the baryon-to-matter ratio in clusters. It is
found that, roughly, $\Omega_{\rm m}\approx 1/3$, which shows that
most of the matter in the universe is non-baryonic i.e. dark
matter.
Moreover, we see that about $2/3$ of the energy density of the
universe is not even in the form of matter and we call it dark
energy. All these results are now confirmed and refined by the
WMAP three year measurements \cite{wmap3} which, combined with
the 2dF galaxy redshift survey \cite{2dFGRS}, yield $\Omega_Bh^2=
0.02223^{+0.00066}_{-0.00083}$, $\Omega_{\rm m}h^2=
0.1262^{+0.0045}_{-0.0062}$, $\Omega_{\rm m}=0.236^{+0.016}_{-0.024}$,
and, assuming that the dark energy is due to a non-zero
cosmological constant, $\Omega_0=0.985^{+0.020}_{-0.016}$.
\vspace{.25cm}
\item[{\bf iii.}] The deceleration parameter
\begin{equation}
q=-\frac{(\ddot{a}/\dot{a})}{(\dot{a}/a)}
=\frac{\rho+3p}{2\rho_{c}}.
\label{eq:decel}
\end{equation}
Measurements of type Ia supernovae \cite{lambda} indicate that the
universe is speeding up ($q_0<0$). This requires that, at present,
$p<0$ as can be seen from Eq.~(\ref{eq:decel}). Negative pressure
can only be attributed to the dark energy since matter is
pressureless. Equation~(\ref{eq:decel}) gives
$q_0=(\Omega_0+3w_X\Omega_X)/2$, where $\Omega_X=\rho_X/\rho_c$
and $w_X=p_X/\rho_X$ with $\rho_X$ and $p_X$ being the dark energy
density and pressure. Observations prefer $w_X=-1$. Actually, the
95\% confidence level limit $w_X<-0.6$ from the Ia supernovae data
combined with constraints from large-scale structure (see
Ref.~\cite{w}) is now improved, after the WMAP three year results
\cite{wmap3} combined with the supernova legacy survey data
\cite{snls}, to $w_X<-0.83$ for a flat universe. Thus,
dark energy can be interpreted as something very close to a
non-zero cosmological constant (see below).
\end{list}

\subsection{Particle Horizon}
\label{subsec:parhor}

\par
Light travels only a finite distance from the time of big bang
($t=0$) until some cosmic time $t$. From Eq.~(\ref{eq:rw}), we
find that the propagation of light along the radial direction is
described by $a(t)dr=dt$. The particle horizon, which is the
instantaneous distance at $t$ travelled by light since $t=0$, is
then
\begin{equation}
d_{H}(t)=a(t)\int_{0}^{t} \frac{dt^{\prime}}{a(t^{\prime})}.
\label{eq:hor}
\end{equation}
The particle horizon is an important notion since it coincides
with the size of the universe already seen at time $t$ or,
equivalently, with the distance over which causal contact has
been established at $t$. Equations~(\ref{eq:expan}) and
(\ref{eq:hor}) give
\begin{equation}
d_{H}(t)=\frac{3\gamma}{3\gamma-2}t,~\gamma\neq \frac{2}{3}.
\label{eq:hort}
\end{equation}
Also,
\begin{equation}
H(t)=\frac{2}{3\gamma}t^{-1},
~d_{H}(t)=\frac{2}{3\gamma-2}H^{-1}(t).
\label{eq:hubblet}
\end{equation}
For matter (radiation), these formulae become
$d_{H}(t)=2H^{-1}(t)=3t$ ($d_{H}(t)=H^{-1}(t)=2t$). Our universe
was matter dominated until fairly recently. So assuming
matter dominance, we can obtain a crude estimate of the present
particle horizon (cosmic time), which is $d_{H}(t_{0})=2H_{0}^{-1}
\approx 6,000~h^{-1}~{\rm{Mpc}}$ ($t_{0}= 2H_{0}^{-1}/3\approx 6.5
\times 10^9~h^{-1}~{\rm years}\approx 9\times 10^9~{\rm years}$).
This is certainly an underestimate and had become a bit of a
problem as independent estimates had suggested longer lifetimes
for some old objects in our universe. After the WMAP three year
measurements \cite{wmap3}, the present age of our universe is
estimated to be $t_{0}=13.73^{+0.13}_{-0.17}\times
10^9~{\rm years}$. The present
$\rho_{c}=3H_{0}^{2}/ 8\pi G\approx 1.9\times 10^{-29}~h^2~
{\rm{gm/cm^{3}}}$.

\subsection{Brief History of the Early Universe}
\label{subsec:history}

\par
We will now briefly summarize the early universe evolution
according to GUTs \cite{ggps}. We will consider a GUT gauge
group $G$ ($={\rm SU}(5)$, ${\rm SO}(10)$, ${\rm SU}(3)^{3}$,
...) with or without SUSY. At a scale $M_{X}\sim 10^{16}
~{\rm{GeV}}$ (the GUT mass scale), $G$
breaks to the standard model gauge group $G_{\rm S}={\rm SU}(3)_c
\times{\rm SU}(2)_{\rm L}\times {\rm U}(1)_Y$ by the VEV of an
appropriate Higgs field $\phi$. (For simplicity, we take this
breaking occurring in one step.) $G_{\rm S}$ is, subsequently,
broken to ${\rm SU}(3)_{c}\times {\rm U}(1)_{\rm em}$ at the
electroweak scale $M_{W}$ (${\rm SU}(3)_{c}$ and
${\rm U}(1)_{\rm em}$ are, respectively, the gauge groups of
strong and electromagnetic interactions).

\par
GUTs together with SBB provide a suitable framework for discussing
the early universe for $t\stackrel{_{>}}{_{\sim }}t_{\rm P}\approx
10^{-44}~{\rm{sec}}$. They predict that the universe, as it
expands and cools, undergoes \cite{kl} a series of phase
transitions during which the gauge symmetry is gradually reduced
and important phenomena take place.

\par
After the big bang, $G$ was unbroken and the universe was filled
with a hot `soup' of massless particles which included photons,
quarks, leptons, gluons, the weak gauge bosons $W^{\pm}$,
$Z^{0}$, the GUT gauge bosons $X$, $Y$, ..., and several Higgs
bosons. In the SUSY case, the SUSY partners were also present.
At $t\sim 10^{-37}~{\rm{sec}}$ ($T\sim 10^{16}~ {\rm{GeV}}$), $G$
broke down to $G_{\rm S}$ and the $X$, $Y$,..., and some Higgs
bosons acquired masses $\sim M_{X}$. Their out-of-equilibrium
decay could, in principle, produce \cite{dimopoulos,bau} the
observed baryon asymmetry of the universe. Important ingredients
are the violation of baryon number, which is inherent in GUTs (as
well as in string inspired models \cite{baryon}), and C and CP
violation. This is the second (potential) success of SBB.

\par
During the GUT phase transition, topologically stable extended
objects \cite{kibble} such as magnetic monopoles \cite{monopole},
cosmic strings \cite{string}, or domain walls \cite{wall} can also
be produced. Monopoles, which exist in most GUTs, can lead into
problems \cite{preskill} which are, however, avoided by inflation
\cite{guth,book,inflation} (see Secs.~\ref{subsec:monopole} and
\ref{subsec:infmono}). This is a period of exponentially fast
expansion of the universe which can occur during some GUT phase
transition and can totally remove the monopoles from the scene.
Alternatively, a more moderate inflation such as thermal inflation
\cite{thermalinf}, which is associated with a phase transition
occurring at a temperature of the order of the electroweak scale,
can dilute them to an acceptable, but possibly measurable level.
Cosmic strings, on the other hand, which are generically present
in many GUT models \cite{cosmicstring,generic}, would contribute
\cite{zel} to the cosmological perturbations which are needed for
structure formation \cite{structure} in the universe leading
\cite{mairi} to extra restrictions on the parameters of the model.
Finally, domain walls are \cite{wall} catastrophic and GUTs should
be constructed so that they avoid them (see e.g.
Ref.~\cite{axion}) or inflation should extinguish them. Note that,
in some cases, more complex extended objects (which are
topologically unstable) such as domain walls bounded by cosmic
strings \cite{wallsbounded} or open cosmic strings connecting
magnetic monopoles \cite{stringsbounded} can be (temporarily)
produced.

\par
At $t\sim 10^{-10}~{\rm{sec}}$ or $T\sim 100 ~{\rm{GeV}}$, the
electroweak phase transition takes place and $G_{\rm S}$ breaks
to ${\rm SU}(3)_{c}\times {\rm U}(1)_{\rm em}$. The weak gauge
bosons $W^{\pm}$, $Z^{0}$, and the electroweak Higgs fields
acquire masses $\sim M_{W}$. Subsequently, at
$t\sim 10^{-4}~{\rm{sec}}$ or $T\sim 1~{\rm{GeV}}$, color
confinement sets in and the quarks get bounded forming hadrons.

\par
The direct involvement of particle physics essentially ends here
since most of the subsequent phenomena fall into the realm of
other branches of physics. We will, however, sketch some of them
since they are crucial for understanding the earlier stages of the
universe evolution where their origin lies.

\par
At $t\approx 180~{\rm{sec}}$ ($T\approx 1~{\rm{MeV}}$),
nucleosynthesis takes place i.e. protons and neutrons form
nuclei. The abundance of light elements (D, $^{3}{\rm He}$,
$^{4}{\rm He}$, $^{6}{\rm Li}$, and $^{7}{\rm Li}$) depends (see
e.g. Ref.~\cite{peebles}) crucially on the number of light
particles (with mass $\stackrel{_{<}}{_{\sim }}
1~{\rm{MeV}}$) i.e. the number of light neutrinos, $N_{\nu}$,
and $\Omega_{B}h^{2}$. Agreement with observations
\cite{deuterium} is achieved for $N_{\nu}=3$ and
$\Omega_{B}h^{2}\approx 0.020$. This is the third success of SBB
cosmology. Much later, at the so-called equidensity point,
$t_{\rm{eq}}\approx 5\times 10^4~{\rm{years}}$, matter dominates
over radiation.

\par
At $t\approx 200,000~h^{-1} {\rm{years}}$ ($T\approx
3,000~{\rm{K}}$), the decoupling of matter and radiation and the
recombination of atoms occur. After this, radiation evolves as
an independent component of the universe and is detected today as
CMBR with temperature $T_{0}\approx 2.73~{\rm{K}}$. The existence
of the CMBR is the fourth success of SBB. Finally, structure
formation \cite{structure} starts at $t\approx 2\times
10^{8}~{\rm{years}}$.

\section{Shortcomings of Big Bang}
\label{sec:short}

\par
The SBB model has been successful in explaining, among other
things, the Hubble expansion, the existence of the CMBR, and the
abundance of the light elements formed during nucleosynthesis.
Despite its successes, this model had a number of long-standing
shortcomings which we will now summarize:

\subsection{Horizon Problem}
\label{subsec:horizon}

\par
The CMBR which we receive now was emitted at the time of
decoupling of matter and radiation when the cosmic temperature
was $T_{\rm d}\approx 3,000 ~\rm{K}$. The decoupling time,
$t_{\rm d}$, can be estimated from
\begin{equation}
\frac{T_0}{T_{\rm d}}=\frac{2.73~\rm{K}}{3,000~\rm{K}}
=\frac{a(t_{\rm d})}{a(t_0)}=\left(\frac {t_{\rm d}}{t_0}
\right)^\frac{2}{3}.
\label{eq:dec}
\end{equation}
It turns out that $t_{\rm d}\approx 200,000~h^{-1}$ years.

\par
The distance over which the CMBR has travelled since its emission
is
\begin{equation}
a(t_0)\int^{t_{0}}_{t_{\rm d}}\frac{dt^\prime}
{a(t^\prime)}=3t_0\left[1-\left(\frac{t_{\rm d}}{t_0}
\right)^\frac{2}{3}\right]\approx 3t_0\approx
6,000~h^{-1}~\rm{Mpc},
\label{eq:lss}
\end{equation}
which coincides with $d_H(t_0)$. A sphere of radius $d_H(t_0)$
around us is called the last scattering surface since the CMBR
has been emitted from it. The particle horizon at decoupling,
$3t_{\rm d}\approx 0.168~h^{-1}~\rm{Mpc}$, expanded until now to
become $0.168~h^{-1} (a(t_0)/a(t_{\rm d}))~{\rm{Mpc}}\approx
184~h^{-1}~\rm{Mpc}$. The angle subtended by this decoupling
horizon now is $\vartheta_{\rm d}\approx 184/6,000\approx 0.03~
\rm{rads}$. Thus, the sky splits into $4\pi/(0.03)^2\approx
14,000$ patches which had never communicated before emitting the
CMBR. The puzzle then is how can the temperature of the black
body radiation from these patches be so finely tuned as COBE
\cite{cobe} and WMAP \cite{wmap1,wmap3} require.

\subsection{Flatness Problem}
\label{subsec:flatness}

\par
The present energy density of the universe has been observed
\cite{dasi} to be very close to its critical value corresponding
to a flat universe ($\Omega_{0}=1.04\pm 0.08$). From
Eq.~(\ref{eq:friedmann}), we obtain $(\rho-\rho_c) /\rho_c=3(8\pi
G\rho_c)^{-1}(k/a^2)\propto a$ for matter. Thus, in the early
universe, $|(\rho-\rho_c)/\rho_c|\ll 1$ and the question is why
the initial energy density of the universe was so finely tuned to
its critical value.

\subsection{Magnetic Monopole Problem}
\label {subsec:monopole}

\par
This problem arises only if we combine SBB with GUTs \cite{ggps}
which predict the existence of magnetic monopoles. According to
GUTs, the universe underwent \cite{kl} a (second order) phase
transition during which an appropriate Higgs field, $\phi$,
developed a non-zero VEV and the GUT gauge group, $G$, broke to
$G_{\rm S}$.

\par
The GUT phase transition produces magnetic monopoles
\cite{monopole}. They are localized deviations from the vacuum
with radius $\sim M_X^{-1}$ and mass $m_M\sim M_X/\alpha_G$
($\alpha_G= g^2_{G}/4\pi$, where $g_G$ is the GUT gauge coupling
constant). The value of $\phi$ on a sphere, $S^2$, of radius
$\gg M_{X}^{-1}$ around the monopole lies on the vacuum manifold
$G/G_{\rm S}$ and we, thus, obtain a mapping: $S^2\rightarrow
G/G_{\rm S}$. If this mapping is homotopically non-trivial, the
monopole is topologically stable.

\par
The initial relative monopole number density must satisfy the
causality bound \cite{einhorn} $r_{M,\rm{in}}\equiv
(n_M/T^3)_{\rm{in}}\stackrel{_{>}}{_{\sim }} \rm{10^{-10}}$
($n_M$ is the monopole number density), which comes from the
requirement that, at monopole production, $\phi$ cannot be
correlated at distances bigger than the particle horizon. The
subsequent evolution of monopoles is studied in
Ref.~\cite{preskill}. The result is that, if $r_{M,\rm{in}}
\stackrel{_{>}}{_{\sim}}\rm{10^{-9}}$ ($\stackrel{_{<}}{_{\sim }}
\rm{10^{-9}}$), the final relative monopole number density
$r_{M,\rm{fin}}\sim 10^{-9}$ ($\sim r_{M,\rm{in}}$). This combined
with the causality bound yields $r_{M,\rm{fin}} \stackrel{_{>}}
{_{\sim }}\rm{10^{-10}}$. However, the requirement that monopoles
do not dominate the energy density of the universe at
nucleosynthesis gives
\begin{equation}
r_M (T \approx 1~\rm{MeV}) \stackrel{_{<}} {_{\sim}}\rm{10^{-19}}
\label{eq:nucleo}
\end{equation}
and we obtain a clear discrepancy of about nine orders of
magnitude.

\subsection{Density Perturbations}
\label{subsec:fluct}

\par
For structure formation \cite{structure} in the universe, we need
a primordial density perturbation, $\delta \rho/\rho$, at all
length scales with a nearly flat spectrum \cite{hz}. We also need
an explanation of the temperature fluctuations of the CMBR
observed by COBE \cite{cobe} and WMAP \cite{wmap1,wmap3} at angles
$\vartheta\stackrel{_{>}}{_{\sim }}\vartheta_{\rm d}\approx 2^{o}$
which violate causality (see Sec.~\ref{subsec:horizon}).

\section{Inflation}
\label{sec:inflation}

\par
The above four cosmological puzzles are solved by inflation
\cite{guth,book,inflation}. Take a real scalar field $\phi$ (the
inflaton) with (symmetric) potential energy density $V(\phi)$
which is quite flat near $\phi=0$ and has minima at
$\phi=\pm\langle\phi\rangle$ with $V(\pm\langle\phi\rangle)=0$. At
high cosmic temperatures, $\phi=0$ due to the temperature
corrections to $V(\phi)$. As the temperature drops, the effective
potential tends to the zero-temperature potential but a small
barrier separating the local minimum at $\phi=0$ and the vacua at
$\phi=\pm\langle\phi\rangle$ remains. At some point, $\phi$
tunnels out to $\phi_1\ll\langle\phi\rangle$ and a bubble with
$\phi=\phi_1$ is created in the universe. The field then rolls
over to the minimum of $V(\phi)$ very slowly (due to the flatness
of the potential $V(\phi)$) with the energy density $\rho\approx
V(\phi=0)\equiv V_0$ remaining practically constant for quite some
time. The Lagrangian density
\begin{equation}
L=\frac{1}{2}\partial_{\mu}\phi
\partial^{\mu}\phi -V(\phi)
\label{eq:lagrange}
\end{equation}
gives the energy-momentum tensor
\begin{equation}
T_{\mu}^{~\nu}=-\partial_\mu\phi\partial^\nu
\phi+\delta_{\mu}^{~\nu}\left(\frac{1}{2}
\partial_\lambda\phi\partial^\lambda\phi-
V(\phi)\right),
\label{eq:energymom}
\end{equation}
which during the slow roll-over becomes $T_{\mu}^{~\nu}\approx
-V_{0}~\delta_{\mu}^{~\nu}$ yielding $\rho\approx -p\approx V_0$.
So, the pressure is opposite to the energy density in accord
with Eq.~(\ref{eq:continuity}). The scale factor $a(t)$ grows
(see below) and the curvature term, $k/a^2$, in
Eq.~(\ref{eq:friedmann}) diminishes. We thus get
\begin{equation}
H^2\equiv\left(\frac{\dot{a}}{a}\right)^2= \frac{8\pi G}{3}V_0,
\label{eq:inf}
\end{equation}
which gives $a(t)\propto e^{Ht}$, $H^2= (8\pi G/3)V_0={\rm
constant}$. So the bubble expands exponentially for some time and
$a(t)$ grows by a factor
\begin{equation}
\frac {a(t_f)}{a(t_i)}={\rm{exp}}H(t_f-t_i)
\equiv{\rm{exp}}H\tau
\label{eq:efold}
\end{equation}
between an initial ($t_i$) and a final ($t_f$) time.

\par
The scenario just described is known as new \cite{new}
inflation. Alternatively, we can imagine, at $t_{\rm P}$, a region of
size $\ell_{\rm P}\sim M_{\rm P}^{-1}$ (the Planck length) where the
inflaton is large and almost uniform carrying negligible kinetic
energy. This region can inflate (exponentially expand) as $\phi$
slowly rolls down towards the vacuum. This scenario is called
chaotic \cite{chaotic} inflation.

\par
We will now show that, with an adequate number of e-foldings,
$N=H\tau$, the first three cosmological puzzles are easily
resolved (we leave the question of density perturbations for
later).

\subsection{Resolution of the Horizon Problem}
\label{subsec:infhor}

\par
The particle horizon during inflation
\begin{equation}
d_H(t)=e^{Ht}\int^t_{t_{i}} \frac{d
t^\prime}{e^{Ht^\prime}}\approx H^{-1}{\rm{exp}}H(t-t_i),
\label{eq:horizon}
\end{equation}
for $t-t_i\gg H^{-1}$, grows as fast as $a(t)$. At $t_f$,
$d_H(t_f)\approx H^{-1}{\rm{exp}}H\tau$ and $\phi$ starts
oscillating about the vacuum. It then decays and reheats
\cite{reheat} the universe at a temperature $T_r\sim
10^9~{\rm{GeV}}$ \cite{gravitino} after which normal big bang
cosmology is recovered. The particle horizon at the end of
inflation $d_H(t_{f})$ is stretched during the
$\phi$-oscillations by a factor $\sim 10^9$ and between $T_r$
and the present time by a factor $T_r/T_0$. So, it finally
becomes equal to $H^{-1}e^{H\tau}10^9(T_r/T_0)$, which must
exceed $2H_{0}^{-1}$ if the horizon problem is to be solved.
This readily holds for $V_0\approx M_{X}^{4}$,
$M_{X}\sim 10^{16}~{\rm GeV}$, and
$N\stackrel{_{>}}{_{\sim }}55$.

\subsection{Resolution of the Flatness Problem}
\label{subsec:infflat}

\par
The curvature term of the Friedmann equation, at present, is
given by
\begin{equation}
\left(\frac{k}{a^2}\right)_0\approx\left(\frac{k}{a^2}
\right)_{\rm bi}e^{-2H\tau}~10^{-18}\left(\frac {10^{-13}~
{\rm{GeV}}}{10^9~{\rm{GeV}}}\right)^2,
\label{eq:curvature}
\end{equation}
where the terms in the RHS are the curvature term before
inflation and its growth factors during inflation,
$\phi$-oscillations, and after reheating. Assuming
$(k/a^2)_{\rm bi}\sim H ^2$, we get $\Omega_0-1
=k/a_{0}^{2}H_{0}^{2}\sim 10^{48}~e^{-2H \tau}\ll 1$ for $H\tau\gg
55$. Strong inflation implies that the present universe is flat
with a great accuracy.

\subsection{Resolution of the Monopole Problem}
\label{subsec:infmono}

\par
For $N\stackrel{_{>}}{_{\sim }} 55$, the magnetic monopoles are
diluted by at least 70 orders of magnitude and become irrelevant.
Also, since $T_r \ll m_M$, there is no magnetic monopole
production after reheating. Extinction of monopoles may also be
achieved by non-inflationary mechanisms such as magnetic
confinement \cite{fate}. For models leading to a possibly
measurable magnetic monopole density see e.g.
Refs.~\cite{thermal,trinification}.

\section{Detailed Analysis of Inflation}
\label{sec:detail}

The Hubble parameter during inflation depends on the value of
$\phi$:
\begin{equation}
H^{2}(\phi)=\frac{8\pi G}{3}V(\phi).
\label{eq:hubble}
\end{equation}
To find the evolution equation for $\phi$ during inflation, we
vary the action
\begin{equation}
\int\sqrt{-{\rm{det}}(g)}~d^{4}x\left(\frac{1}{2}
\partial_ {\mu}\phi\partial^{\mu}\phi-V(\phi)+
M(\phi)\right),
\label{eq:action}
\end{equation}
where $g$ is the metric tensor and $M(\phi)$ the (trilinear)
coupling of $\phi$ to light matter causing its decay. Assuming
that this coupling is weak, one finds \cite{dh}
\begin{equation}
\ddot{\phi}+3H\dot{\phi}+\Gamma_{\phi}\dot{\phi}+
V^{\prime}(\phi)=0,
\label{eq:evolution}
\end{equation}
where the prime denotes derivation with respect to $\phi$ and
$\Gamma_{\phi}$ is the decay width \cite{width} of the inflaton.
Assume, for the moment, that the decay time of $\phi$, $t_d=
\Gamma_{\phi}^{-1}$, is much greater than $H^{-1}$, the expansion
time for inflation. Then the term $\Gamma_{\phi}\dot{\phi}$ can be
ignored and Eq.~(\ref{eq:evolution}) becomes
\begin{equation}
\ddot{\phi}+3H\dot{\phi}+V^{\prime}(\phi)=0.
\label{eq:reduce}
\end{equation}
Inflation is by definition the situation where $\ddot{\phi}$ is
subdominant to the `friction term' $3H\dot{\phi}$ (and the kinetic
energy density is subdominant to the potential one).
Equation~(\ref{eq:reduce}) then reduces to the inflationary
equation
\cite{slowroll}
\begin{equation}
3H\dot{\phi}=-V^{\prime}(\phi),
\label{eq:infeq}
\end{equation}
which gives
\begin{equation}
\ddot{\phi}=-\frac{V^{\prime\prime}(\phi)\dot{\phi}}
{3H(\phi)}+\frac{V^{\prime}(\phi)}
{3H^{2}(\phi)}H^\prime(\phi)\dot{\phi}.
\label{eq:phidd}
\end{equation}
Comparing the two terms in the RHS of this equation with the
friction term in Eq.~(\ref{eq:reduce}), we get the conditions
for inflation (slow roll conditions):
\begin{equation}
\epsilon,~|\eta|\leq 1~~{\rm with}~~\epsilon\equiv
\frac{M_{\rm P}^{2}}{16\pi}\left(\frac{V^{\prime}(\phi)}
{V(\phi)}\right)^{2},~\eta\equiv\frac{M_{\rm P}^{2}}
{8\pi}\frac{V^{\prime\prime}(\phi)}{V(\phi)}.
\label{eq:src}
\end{equation}
The end of the slow roll-over occurs when either of these
inequalities is saturated. If $\phi_f$ is the value of $\phi$ at
the end of inflation, then $t_f \sim H^{-1}(\phi_f)$.

\par
The number of e-foldings during inflation can be calculated as
follows:
\begin{equation}
N(\phi_{i}\rightarrow \phi_{f})\equiv\ln \left(\frac{a(t_{f})}
{a(t_{i})}\right)=\int^{t_{f}} _{t_{i}} Hdt=
\int^{\phi_{f}}_{\phi_{i}}\frac{H (\phi)}
{\dot{\phi}}d\phi=-\int^{\phi_{f}}_{\phi_{i}} \frac {3 H^2 (\phi)
d \phi}{V^{\prime}(\phi)},
\label{eq:nefolds}
\end{equation}
where Eqs.~(\ref{eq:efold}), (\ref{eq:infeq}) were used. We shift
$\phi$ so that the global minimum of $V(\phi)$ is displaced at
$\phi$=0. Then, if $V(\phi)=\lambda \phi^{\nu}$ during inflation,
we have
\begin{equation}
N(\phi_{i} \rightarrow \phi_{f})= -\int^{\phi_{f}}_{\phi_{i}}
\frac {3H^2(\phi)d\phi}{V^{\prime}(\phi)}= -8\pi
G\int^{\phi_{f}}_{\phi_{i}} \frac
{V(\phi)d\phi}{V^{\prime}(\phi)}= \frac {4 \pi
G}{\nu}(\phi^{2}_{i}-\phi^{2}_{f}).
\label{eq:expefold}
\end{equation}
Assuming that $\phi_{i}\gg\phi_{f}$, this reduces to
$N(\phi)\approx(4 \pi G/\nu)\phi^2$.

\section{Coherent Oscillations of the Inflaton}
\label{sec:osci}

After the end of inflation at $t_f$, the term $\ddot{\phi}$ takes
over in Eq.~(\ref{eq:reduce}) and $\phi$ starts performing coherent
damped oscillations about the global minimum of the potential. The
rate of energy density loss, due to friction, is given by
\begin{equation}
\dot{\rho}=\frac{d}{dt}\left(\frac{1}{2}
\dot{\phi}^2+V(\phi)\right)=-3H\dot{\phi}^2= -3H(\rho+p),
\label{eq:damp}
\end{equation}
where $\rho=\dot{\phi}^2/2+V(\phi)$ and
$p=\dot{\phi}^2/2-V(\phi)$. Averaging $p$ over one oscillation of
$\phi$ and writing $\rho+p= \gamma\rho$, we get $\rho\propto
a^{-3\gamma}$ and $a(t)\propto t^{2/3\gamma}$ (see
Sec.~\ref{subsec:friedmann}).

\par
The number $\gamma$ can be written as (assuming a symmetric
potential)
\begin{equation}
\gamma=\frac{\int^{T}_{0}\dot{\phi}^{2}dt} {\int^{T}_{0}\rho dt}=
\frac{\int^{\phi_{{\rm{max}}}}_{0}
\dot{\phi}d\phi}{\int^{\phi_{{\rm{max}}}}_{0}
(\rho/\dot{\phi})d\phi},
\label{eq:gamma}
\end{equation}
where $T$ and $\phi_{{\rm{max}}}$ are the period and the amplitude
of the oscillation. From
$\rho=\dot{\phi}^2/2+V(\phi)=V_{{\rm{max}}}$, where $V_{\rm max}$
is the maximal potential energy density, we obtain $\dot{\phi}=
\sqrt{2(V_{{\rm{max}}}-V(\phi))}$. Substituting this in
Eq.~(\ref{eq:gamma}), we get \cite{oscillation}
\begin{equation}
\gamma=\frac{2\int^{\phi_{{\rm{max}}}}_{0}
(1-V/V_{{\rm{max}}})^\frac{1}{2} d\phi}
{\int^{\phi_{{\rm{max}}}}_{0}
(1-V/V_{{\rm{max}}})^{-\frac{1}{2}}d\phi}.
\label{eq:gammafinal}
\end{equation}
For $V(\phi)=\lambda\phi^{\nu}$, we find $\gamma= 2\nu/(\nu+2)$
and, thus, $\rho\propto a^{-6\nu/(\nu+2)}$ and $a(t)\propto
t^{(\nu+2)/3\nu}$. For $\nu=2$, in particular, $\gamma=1$,
$\rho\propto a^{-3}$, $a(t)\propto t^{2/3}$ and $\phi$ behaves
like pressureless matter. This is not unexpected since a coherent
oscillating massive free field corresponds to a distribution of
static massive particles. For $\nu$=4, we obtain $\gamma=4/3$,
$\rho\propto a^{-4}$, $a(t)\propto t^{1/2}$ and the system
resembles radiation. For $\nu = 6$, one has $\gamma=3/2$,
$\rho\propto a^{-9/2}$, $a(t) \propto t^{4/9}$ and the expansion
is slower (the pressure is higher) than in radiation.

\section{Decay of the Inflaton}
\label{sec:decay}

Reintroducing the decay term $\Gamma_{\phi} \dot{\phi}$,
Eq.~(\ref{eq:evolution}) can be written as
\begin{equation}
\dot{\rho}=\frac{d}{dt}\left(\frac{1}{2}\dot{\phi}^2
+V(\phi)\right)=-(3H+\Gamma_\phi)\dot{\phi}^2,
\label{eq:decay}
\end{equation}
which is solved \cite{reheat,oscillation} by
\begin{equation}
\rho(t)=\rho_{f} \left(\frac{a(t)}{a(t_{f})} \right)^{-3
\gamma}{\rm{exp}}[-\gamma \Gamma_{\phi}(t-t_f)],
\label{eq:rho}
\end{equation}
where $\rho_f$ is the energy density at $t_f$. The second and
third factors in the RHS of this equation represent the dilution
of the field energy due to the expansion of the universe and the
decay of $\phi$ to light particles respectively.

\par
All pre-existing radiation (known as old radiation) was diluted
by inflation, so the only radiation present is the one produced by
the decay of $\phi$ and is known as new radiation. Its energy
density $\rho_r$ satisfies \cite{reheat,oscillation} the equation
\begin{equation}
\dot{\rho}_{r}=-4H \rho_{r}+\gamma\Gamma_{\phi}\rho,
\label{eq:newrad}
\end{equation}
where the first term in the RHS represents the dilution of
radiation due to the cosmological expansion while the second one
is the energy density transfer from $\phi$ to radiation. Taking
$\rho_{r}(t_f)$=0, this equation gives \cite{reheat,oscillation}
\begin{equation}
\rho_{r}(t)=\rho_{f}\left(\frac {a(t)}
{a(t_{f})}\right)^{-4}\int^{t}_{t_{f}} \left(\frac{a(t^{\prime})}
{a(t_{f})}\right)^{4-3 \gamma} e^{ -\gamma \Gamma_{\phi}
(t^{\prime}-t_f)} ~\gamma\Gamma_{\phi} dt^{\prime}.
\label{eq:rad}
\end{equation}
For $t_{f} \ll t_{d}$ and $\nu =2$, this expression is
approximated by
\begin{equation}
\rho_{r}(t)=\rho_{f}\left(\frac {t}{t_f}
\right)^{-\frac{8}{3}}\int^{t}_{0}\left(
\frac{t^{\prime}}{t_{f}}\right)^\frac{2}{3}
e^{-\Gamma_{\phi}t^{\prime}}dt^{\prime},
\label{eq:appr}
\end{equation}
which can be expanded as
\begin{equation}
\rho_{r}=\frac {3}{5}~\rho~\Gamma_{\phi}t\left[1+
\frac{3}{8}~\Gamma_{\phi}t+\frac
{9}{88}~(\Gamma_{\phi}t)^2+\cdots\right]
\label{eq:expand}
\end{equation}
with $\rho=\rho_{f} (t/t_{f})^{-2}{\rm{exp}} (-\Gamma_{\phi}t)$
being the energy density of the field $\phi$.

\par
The energy density of the new radiation grows relative to the
energy density of the oscillating field and becomes essentially
equal to it at a cosmic time $t_{d}=\Gamma_{\phi}^{-1}$ as one can
deduce from Eq.~(\ref{eq:expand}). After this time, the universe
enters into the radiation dominated era and the normal big bang
cosmology is recovered. The temperature at $t_{d}$,
$T_{r}(t_{d})$, is historically called the reheat temperature
although no supercooling and subsequent reheating of the universe
actually takes place. Using Eq.~(\ref{eq:temptime}), we find that
\begin{equation}
T_{r}=\left(\frac {45}{16 \pi^{3}g_*}
\right)^\frac{1}{4}(\Gamma_{\phi}M_{\rm P})^\frac{1}{2},
\label{eq:reheat}
\end{equation}
where $g_*$ is the effective number of massless degrees of
freedom. For $V(\phi)=\lambda\phi^{\nu}$, the total expansion of
the universe during the damped field oscillations is
\begin{equation}
\frac{a(t_{d})}{a(t_{f})}=\left(
\frac{t_{d}}{t_{f}}\right)^{\frac{\nu+2}{3\nu}}.
\label{eq:expansion}
\end{equation}

\section{Density Perturbations from Inflation}
\label{sec:density}

Inflation not only homogenizes the universe but also generates the
density perturbations needed for structure formation. To see this,
we introduce the notion of event horizon at $t$. This includes all
points with which we will eventually communicate sending signals
at $t$. Its instantaneous radius is
\begin{equation}
d_{e}(t)=a(t)\int^{\infty}_{t}
\frac{dt^{\prime}}{a(t^{\prime})}.
\label{eq:event}
\end{equation}
This yields an infinite event horizon for matter or radiation.
For inflation, however, we obtain $d_{e}(t)=H^{-1} <\infty$, which
varies slowly with $t$. Points in our event horizon at $t$ with
which we can communicate sending signals at $t$ are eventually
pulled away by the exponential expansion and we cease to be able
to communicate with them emitting signals at later times. We say
that these points (and the corresponding scales) crossed outside
the event horizon. Actually, the exponentially expanding (de
Sitter) space is like a black hole turned inside out. Then,
exactly as in a black hole, there are quantum fluctuations of the
thermal type governed by the Hawking temperature
\cite{hawking,gibbons} $T_{H}=H/2\pi$. It turns out
\cite{bunch,vilenkin} that the quantum fluctuations of all
massless fields (the inflaton is nearly massless due to the
flatness of the potential) are $\delta\phi=T_{H}$. These
fluctuations of $\phi$ lead to energy density perturbations
$\delta\rho= V^{\prime}(\phi)\delta\phi$. As the scale of this
perturbations crosses outside the event horizon, they become
\cite{fischler} classical metric perturbations.

\par
It has been shown \cite{zeta} (for a review, see e.g.
Ref.~\cite{zetarev}) that the quantity
$\zeta\approx\delta\rho/(\rho+p)$ remains constant outside the
event horizon. Thus, the density perturbation at any present
physical (comoving) scale $\ell$, $(\delta\rho/\rho)_{\ell}$,
when this scale crosses inside the post-inflationary particle
horizon ($p$=0 at this instance) can be related to the value of
$\zeta$ when the same scale crossed outside the inflationary event
horizon (at $\ell\sim H^{-1}$). This latter value of $\zeta$ is
found, using Eq.~(\ref{eq:infeq}), to be
\begin{equation}
\zeta\mid_{\ell\sim H^{-1}}=\left(\frac
{\delta\rho}{\dot{\phi}^2}\right)_{\ell\sim
H^{-1}}=\left(\frac{V^{\prime}(\phi)H(\phi)}{2\pi
\dot{\phi}^2}\right)_{\ell\sim H^{-1}}=-\left( \frac {9
H^{3}(\phi)} {2\pi V^{\prime}(\phi)} \right)_{\ell\sim
H^{-1}}.
\label{eq:zeta}
\end{equation}
Taking into account an extra 2/5 factor from the fact that the
universe is matter dominated when the scale $\ell$ re-enters the
horizon, we obtain
\begin{equation}
\left(\frac{\delta\rho}{\rho}\right)_{\ell}=
\frac{16\sqrt{6\pi}}{5}~\frac {V^\frac{3}{2}
(\phi_{\ell})}{M^{3}_{\rm P} V^{\prime}(\phi_{\ell})}.
\label{eq:deltarho}
\end{equation}

\par
The calculation of $\phi_{\ell}$, the value of the field $\phi$
when the comoving scale $\ell$ crossed outside the event horizon,
goes as follows. At the reheat temperature $T_r$, a comoving
(present physical) scale $\ell$ was equal to
$\ell(a(t_{d})/a(t_{0}))=\ell(T_{0}/T_{r})$. Its
magnitude at $t_{f}$ was equal to
$\ell(T_{0}/T_{r})(a(t_{f})/a(t_{d}))=
\ell(T_{0}/T_{r})(t_{f}/t_{d})^{(\nu+2)/3 \nu}
\equiv\ell_{{\rm{phys}}}(t_{f})$, where the potential
$V(\phi)=\lambda\phi^{\nu}$ was assumed. The scale $\ell$, when it
crossed outside the inflationary event horizon, was equal to
$H^{-1}(\phi_{\ell})$. We, thus, obtain
\begin{equation}
H^{-1}(\phi_{\ell}) e^{N(\phi_{\ell})} =
\ell_{{\rm{phys}}}(t_{f}),
\label{eq:lphys}
\end{equation}
which gives $\phi_{\ell}$ and, thus, $N(\phi_{\ell}) \equiv
N_{\ell}$, the number of e-foldings the comoving scale $\ell$
suffered during inflation. In particular, the number of
e-foldings suffered by our present horizon $\ell=2H_{0}^{-1}
\sim 10^4~{\rm Mpc}$ turns out to be $N_{Q}\approx 50-60$.

\par
Taking $V(\phi)=\lambda \phi^4$, Eqs.~(\ref{eq:expefold}),
(\ref{eq:deltarho}), and (\ref{eq:lphys}) give
\begin{equation}
\left(\frac{\delta\rho}{\rho}\right)_{\ell}=
\frac{4\sqrt{6\pi}}{5}\lambda^\frac{1}{2}\left(
\frac{\phi_{\ell}}{M_{\rm P}}\right)^3= \frac{4 \sqrt{6
\pi}}{5}\lambda^\frac{1}{2}
\left(\frac{N_{\ell}}{\pi}\right)^\frac{3}{2}.
\label{eq:nl}
\end{equation}
From the result of COBE \cite{cobe}, $(\delta\rho/\rho)_{Q}\approx
6\times 10^{-5}$, one can then deduce that $\lambda\approx 6\times
10^{-14}$ for $N_{Q}\approx 55$. We thus see that the inflaton
must be a very weakly coupled field. In non-SUSY GUTs, the
inflaton is necessarily gauge singlet since otherwise radiative
corrections will make it strongly coupled. This is not so
satisfactory since it forces us to introduce an otherwise
unmotivated very weakly coupled gauge singlet. In SUSY GUTs,
however, the inflaton could be identified \cite{nonsinglet} with a
conjugate pair of gauge non-singlet fields $\bar{\phi}$, $\phi$
already present in the theory and causing the gauge symmetry
breaking. Absence of strong radiative corrections from gauge
interactions is guaranteed by the mutual cancellation of the
D-terms of these fields.

\par
The spectrum of density perturbations can be analyzed. For
$V(\phi)=\lambda\phi^{\nu}$, we find
$(\delta\rho/\rho)_{\ell}\propto \phi_{\ell}^{(\nu+2)/2}$ which,
together with $N(\phi_{\ell})\propto\phi_{\ell}^{2}$ (see
Eq.~(\ref{eq:expefold})), gives
\begin{equation}
\left(\frac{\delta\rho}{\rho}\right)_{\ell}=\left(
\frac{\delta\rho}{\rho}\right)_{Q}\left(
\frac{N_{\ell}}{N_{Q}}\right)^{\frac{\nu+2}{4}}.
\label{eq:spectrum}
\end{equation}
The scale $\ell$ divided by the size of our present horizon
($2H_{0}^{-1}\sim 10^4~{\rm{Mpc}}$) should equal ${\rm
exp}(N_{\ell}-N_{Q})$. This gives
$N_{\ell}/N_{Q}=1+\ln(\ell/2H_{0}^{-1})^{1/N_{Q}}$ which expanded
around $\ell=2H_{0}^{-1}$ and substituted in Eq.~(\ref{eq:spectrum})
yields
\begin{equation}
\left(\frac{\delta \rho}{\rho}\right)_{\ell}\approx
\left(\frac{\delta \rho}{\rho}\right)_{Q}\left(
\frac{\ell}{2H_{0}^{-1}}\right)^{\alpha_{s}}
\label{eq:alphas}
\end{equation}
with $\alpha_{s}=(\nu+2)/4N_{Q}$. For $\nu=4$, $\alpha_{s}\approx
0.03$ and, thus, the density perturbations are essentially scale
independent. The customarily used spectral index
$n_s=1-2\alpha_{s}$ is about 0.94 in this case.

\section{Temperature Fluctuations}
\label{sec:temperature}

The density inhomogeneities produce temperature fluctuations in
the CMBR. For angles $\vartheta \stackrel{_{>}}{_{\sim }}2^{o}$,
the dominant effect is the scalar Sachs-Wolfe \cite{sachswolfe}
effect. Density perturbations on the last scattering surface
cause scalar gravitational potential fluctuations, which then
produce temperature fluctuations in the CMBR. The reason is that
regions with a deep gravitational potential will cause the photons
to lose energy as they climb up the potential well and, thus, these
regions will appear cooler.

\par
Analyzing the temperature fluctuations from the scalar Sachs-Wolfe
effect in spherical harmonics, we obtain the corresponding
quadrupole anisotropy:
\begin{equation}
\left(\frac{\delta T}{T}\right)_{Q-S}=\left( \frac{32
\pi}{45}\right)^\frac{1}{2}
\frac{V^\frac{3}{2}(\phi_{\ell})}{M^{3}_{\rm P}
V^{\prime}(\phi_{\ell})}.
\label{eq:quadrupole}
\end{equation}
For $V(\phi)=\lambda\phi^{\nu}$, this becomes
\begin{equation}
\left(\frac{\delta T}{T}\right)_{Q-S}=\left(
\frac{32\pi}{45}\right)^\frac{1}{2} \frac{\lambda^\frac{1}{2}
\phi_{\ell}^{\frac{\nu+2}{2}}}{\nu M^{3}_{\rm P}} =\left(\frac{32
\pi}{45}\right)^\frac{1}{2} \frac{\lambda^\frac{1}{2}}{\nu
M^{3}_{\rm P}} \left(\frac{\nu M^{2}_{\rm P}}{4\pi}
\right)^{\frac{\nu+2}{4}} N_{\ell}^{\frac{\nu+2}{4}}.
\label{eq:anisotropy}
\end{equation}
Comparing this with the COBE \cite{cobe} result, $(\delta
T/T)_{Q}\approx 6.6\times 10^{-6}$, we obtain $\lambda\approx
6\times 10^{-14}$ for $\nu=4$ and number of e-foldings suffered by
our present horizon scale during the inflationary phase
$N_{Q}\approx 55$.

\par
There are also tensor fluctuations \cite{tensor} in the
temperature of the CMBR. The tensor quadrupole anisotropy is
\begin{equation}
\left(\frac{\delta T}{T}\right)_{Q-T}
\approx 0.77~\frac
{V^{\frac{1}{2}}(\phi_{\ell})}{M^{2}_{\rm P}}.
\label{eq:tensor}
\end{equation}
The total quadrupole anisotropy is given by
\begin{equation}
\left(\frac{\delta T}{T}\right)_{Q}=
\left[\left(\frac{\delta T}{T}\right)^{2}_{Q-S}+
\left(\frac{\delta T}{T}\right)^{2}_{Q-T}
\right]^{\frac{1}{2}}
\label{eq:total}
\end{equation}
and the ratio
\begin{equation}
r=\frac{\left(\delta T/T\right)^{2}_{Q-T}}
{\left(\delta T/T\right)^{2}_{Q-S}}\approx 0.27
~\left(\frac{M_{\rm P}V^{\prime}(\phi_{\ell})}
{V(\phi_{\ell})}\right)^{2}.
\label{eq:ratio}
\end{equation}
For $V(\phi)=\lambda\phi^{\nu}$, we obtain
$r\approx 3.4~\nu/N_Q\ll 1$, and the tensor
contribution to the temperature fluctuations of the CMBR is
negligible. Actually, the tensor fluctuations turn out to be
negligible in all the cases considered here.

\section{Hybrid Inflation}
\label{sec:hybrid}

\subsection{The non-Supersymmetric Version}
\label{subsec:nonsusy}

The main disadvantage of inflationary scenarios such as the new
\cite{new} or the chaotic \cite{chaotic} scenario is that they
require tiny parameters in order to reproduce the results of COBE
\cite{cobe}. This has led Linde \cite{hybrid} to propose, in the
context of non-SUSY GUTs, hybrid inflation which uses two real
scalar fields $\chi$ and $\sigma$ instead of one. The field $\chi$
provides the vacuum energy density which drives inflation, while
$\sigma$ is the slowly varying field during inflation. This
splitting of roles allows us to reproduce the COBE results with
natural (not too small) values of the parameters.

\par
The scalar potential utilized by Linde is
\begin{equation}
V(\chi,\sigma)=\kappa^2 \left(M^2-\frac{\chi^2}{4}
\right)^2+\frac{\lambda^2\chi^2 \sigma^2}{4}+
\frac{m^2\sigma^2}{2},
\label{eq:lindepot}
\end{equation}
where $\kappa,~\lambda>0$ are dimensionless constants and $M,~m>0$
mass parameters. The vacua lie at $\langle\chi\rangle=\pm 2M$,
$\langle\sigma\rangle=0$. For $m$=0, $V$ has a flat direction at
$\chi=0$, where $V=\kappa^2M^4$ and the ${\rm mass}^2$ of $\chi$
is $m^2_\chi= -\kappa^2M^2+\lambda^2\sigma^2/2$. So, for $\chi=0$
and $\vert\sigma\vert>\sigma_c\equiv \sqrt{2}\kappa M/ \lambda$,
we obtain a flat valley of minima. For $m\neq 0$, the valley
acquires a slope and the system can inflate as the field $\sigma$
slowly rolls down this valley.

\par
The $\epsilon$ and $\eta$ criteria (see Eq.~(\ref{eq:src})) imply
that inflation continues until $\sigma$ reaches $\sigma_c$, where
it terminates abruptly. It is followed by a waterfall i.e. a
sudden entrance into an oscillatory phase about a global minimum.
Since the system can fall into either of the two minima with equal
probability, topological defects (magnetic monopoles, cosmic
strings, or domain walls) are copiously produced \cite{smooth} if
they are predicted by the particular GUT model employed. So, if
the underlying GUT gauge symmetry breaking
(by $\langle\chi\rangle$)
leads to the existence of magnetic monopoles or domain walls, we
encounter a cosmological catastrophe.

\par
The onset of hybrid inflation requires \cite{onset} that, at
$t\sim H^{-1}$, $H$ being the inflationary Hubble parameter, a
region exists with size $\stackrel{_{>}}{_{\sim}}H^{-1}$ where
$\chi$ and $\sigma$ are almost uniform with negligible kinetic
energies and values close to the bottom of the valley of minima.
Such a region, at $t_{\rm P}$, would have been much larger than the
Planck length $\ell_{\rm P}$ and it is, thus, difficult to imagine how
it could be so homogeneous. Moreover, as it has been argued
\cite{initial}, the initial values (at $t_{\rm P}$) of the fields in
this region must be strongly restricted in order to obtain
adequate inflation. Several possible solutions to this problem of
initial conditions for hybrid inflation have been proposed (see
e.g. Refs.~\cite{double,sugra,costas}).

\par
The quadrupole anisotropy of the CMBR produced during hybrid
inflation can be estimated, using Eq.~(\ref{eq:quadrupole}), to
be
\begin{equation}
\left(\frac{\delta T}{T}\right)_{Q}\approx
\left(\frac{16\pi}{45}\right)^{\frac{1}{2}}
\frac{\lambda\kappa^2M^5}{M^3_{\rm P}m^2}.
\label{eq:lindetemp}
\end{equation}
The COBE \cite{cobe} result, $(\delta T/T)_{Q} \approx 6.6\times
10^{-6}$, can then be reproduced with $M\approx 2.86\times 10^{16}
~{\rm GeV}$, the SUSY GUT VEV, and $m\approx 1.3
~\kappa\sqrt{\lambda}\times 10^{15}~{\rm GeV}$. Note that $m\sim
10^{12}~{\rm GeV}$ for $\kappa$, $\lambda\sim 10^{-2}$.

\subsection{The Supersymmetric Version}
\label{subsec:susy}

It has been observed \cite{lyth} that hybrid inflation is  tailor
made for globally SUSY GUTs except that an intermediate scale
mass for $\sigma$ cannot be obtained. Actually, all scalar fields
acquire masses $\sim m_{3/2} \sim 1~{\rm TeV}$ (the gravitino
mass) from soft SUSY breaking.

\par
Let us consider the renormalizable superpotential
\begin{equation}
W=\kappa S(-M^2+\bar{\phi}\phi),
\label{eq:superpot}
\end{equation}
where $\bar{\phi}$, $\phi$ is a  pair of $G_{\rm S}$ singlet left
handed superfields belonging to conjugate representations of $G$
and reducing its rank by their VEVs and $S$ is a gauge singlet
left handed superfield. The parameters $\kappa$ and $M$
($\sim 10^{16}~{\rm GeV}$) are made positive by field
redefinitions. The vanishing of the F-term $F_S$ gives
$\langle\bar{\phi}\rangle\langle\phi\rangle=M^2$ and the D-terms
vanish for $\vert\langle\bar{\phi}
\rangle\vert=\vert\langle\phi\rangle\vert$. So, the SUSY vacua lie
at $\langle\bar{\phi} \rangle^*=\langle\phi\rangle=\pm M$ (after
rotating $\bar{\phi}$, $\phi$ on the real axis by a $G$
transformation) and $\langle S\rangle=0$ (from $F_{\bar{\phi}}=
F_{\phi}=0$). Thus, the superpotential $W$ leads to the breaking
of $G$.

\par
The interesting observation \cite{lyth} is that the same
superpotential $W$ also gives rise to hybrid inflation. The
potential derived from $W$ is
\begin{equation}
V(\bar{\phi},\phi,S)=\kappa^2\vert M^2- \bar{\phi}\phi\vert^2+
\kappa^2\vert S\vert^2(\vert\bar{\phi}\vert^2+
\vert\phi\vert^2)+{\rm D-terms}.
\label{eq:hybpot}
\end{equation}
D-flatness implies that $\bar{\phi}^*=e^{i\theta}\phi$. We take
$\theta=0$, so that the SUSY vacua are contained. The
superpotential $W$ has a ${\rm U}(1)_{\rm R}$ R symmetry
\cite{rsym}: $\bar{\phi}\phi\to\bar{\phi}\phi$,
$S\to e^{i\alpha}S$, $W\to e^{i\alpha}W$. Note, in passing,
that global continuous symmetries such as this R symmetry can
effectively arise \cite{laz1} from the rich discrete symmetry
groups encountered in many compactified string theories (see e.g.
Ref.~\cite{laz2}). It is important to point out that $W$ is the
most general renormalizable superpotential allowed by $G$ and
${\rm U}(1)_{\rm R}$. Bringing the fields $\bar{\phi}$, $\phi$,
and $S$ on the real axis by appropriate $G$ and
${\rm U}(1)_{\rm R}$ transformations, we write $\bar{\phi}=
\phi\equiv\chi/2$ and $S\equiv\sigma/\sqrt{2}$, where $\chi$ and
$\sigma$ are normalized real scalar fields. The potential $V$ in
Eq.~(\ref{eq:hybpot}) then takes the form in Eq.~(\ref{eq:lindepot})
with $\kappa=\lambda$ and $m=0$. So, Linde's potential for hybrid
inflation is almost obtainable from SUSY GUTs, but without the
mass term of $\sigma$.

\par
SUSY breaking by the vacuum energy density $\kappa^2M^4$ on the
inflationary valley ($\bar{\phi}=\phi=0$, $\vert
S\vert>S_{c}\equiv M$) causes a mass splitting in the
supermultiplets $\bar{\phi}$, $\phi$. We obtain a Dirac fermion
with ${\rm mass}^2=\kappa^2\vert S\vert^2$ and two complex scalars
with ${\rm mass}^2=\kappa^2\vert S \vert^2\pm\kappa^2M^2$. This
leads \cite{dss} to one-loop radiative corrections to $V$ on the
valley which are calculated by using the Coleman-Weinberg formula
\cite{cw}:
\begin{equation}
\Delta V=\frac{1}{64\pi^2}\sum_i(-)^{F_i}\ M_i^4
\ln\frac{M_i^2}{\Lambda^2},
\label{eq:deltav}
\end{equation}
where the sum extends over all helicity states $i$ with fermion
number $F_i$ and ${\rm mass}^2=M_i^2$ and $\Lambda$ is a
renormalization scale. We find that
\begin{equation}
\Delta V(\vert S\vert)=\kappa^2 M^4~{\kappa^2\mathcal{N}
\over 32\pi^2}
\left( 2\ln{\kappa^2\vert S\vert^2\over\Lambda^2}
+(z+1)^{2}\ln(1+z^{-1})+(z-1)^{2}\ln(1-z^{-1})\right),
\label{eq:rc}
\end{equation}
where $z\equiv x^2\equiv \vert S\vert^2/M^2$ and $\mathcal{N}$
is the dimensionality of the representations to which
$\bar{\phi}$, $\phi$ belong. These radiative corrections
generate the necessary slope on the inflationary valley. Note
that the slope is $\Lambda$-independent.

\par
From Eqs.~(\ref{eq:expefold}), (\ref{eq:quadrupole}), and
(\ref{eq:rc}), we find the quadrupole anisotropy of the CMBR:
\begin{equation}
\left(\frac{\delta T}{T}\right)_{Q}\approx
\frac{8\pi}{\sqrt{\mathcal{N}}}\left(\frac{N_{Q}}{45}
\right)^{\frac{1}{2}}\left(\frac{M}{M_{\rm P}}
\right)^2x_Q^{-1}y_Q^{-1}\Lambda(x_Q^2)^{-1}
\label{eq:qa}
\end{equation}
with
\begin{equation}
\Lambda(z)=(z+1)\ln(1+z^{-1})+(z-1)\ln(1-z^{-1}),
\label{eq:lambda}
\end{equation}
\begin{equation}
y_Q^2=\int_1^{x_Q^2}\frac{dz}{z}\Lambda(z)^{-1},~y_Q\geq 0.
\label{eq:yq}
\end{equation}
Here, $x_Q\equiv \vert S_Q\vert/M$ with $S_Q$ being the value of
$S$ when our present horizon crossed outside the inflationary
horizon. Finally, from Eq.~(\ref{eq:rc}), one finds
\begin{equation}
\kappa\approx\frac{8\pi^{\frac{3}{2}}}{\sqrt{\mathcal{N}N_Q}}
~y_Q~\frac{M}{M_{\rm P}}.
\label{eq:kappa}
\end{equation}

\par
The slow roll conditions for SUSY hybrid inflation are
$\epsilon,\vert\eta\vert\leq 1$, where
\begin{equation}
\epsilon=\left(\frac{\kappa^2M_{\rm P}}{16\pi^2M}\right)^2
\frac{\mathcal{N}^2x^2}{8\pi}\Lambda(x^2)^2,
\label{eq:epsilon}
\end{equation}
\begin{equation}
\eta=\left(\frac{\kappa M_{\rm P}}{4\pi M}\right)^2
\frac{\mathcal{N}}{8\pi}\left((3z+1)\ln(1+z^{-1})+
(3z-1)\ln(1-z^{-1})\right).
\label{eq:eta}
\end{equation}
These conditions are violated only `infinitesimally' close to the
critical point ($x=1$). So, inflation continues until this point,
where the waterfall occurs.

\par
Using COBE \cite{cobe} and eliminating $x_Q$ between
Eqs.~(\ref{eq:qa}) and (\ref{eq:kappa}), we obtain $M$ as a
function of $\kappa$. The maximal $M$ which can be achieved is
$\approx 10^{16}~{\rm GeV}$ (for $\mathcal{N}=8$,
$N_Q\approx 55)$ and, although somewhat smaller than
the SUSY GUT VEV, is quite close to it. As an example, take
$\kappa=4\times 10^{-3}$ which gives $M\approx 9.57\times
10^{15}~{\rm GeV}$, $x_Q\approx2.633$, $y_Q\approx 2.42$. The slow
roll conditions are violated at $x-1\approx 7.23\times 10^{-5}$,
where $\eta=-1$ ($\epsilon\approx 8.17 \times 10^{-8}$ at $x=1$).
The spectral index $n_s=1-6\epsilon+2\eta$ \cite{liddle} is about
0.985.

\par
SUSY hybrid inflation is considered natural for the following
reasons:
\begin{list}
\setlength{\rightmargin=0cm}{\leftmargin=0cm}
\item[{\bf i.}] There is no need of tiny coupling constants
($\kappa\sim 10^{-3}$).
\vspace{.25cm}
\item[{\bf ii.}] The superpotential $W$ in Eq.~(\ref{eq:superpot})
has the most general renormalizable form allowed by $G$ and
${\rm U}(1)_{\rm R}$. The coexistence of the $S$ and
$S\bar{\phi}\phi$ terms implies that $\bar{\phi}\phi$ is neutral
under all symmetries and, thus, all the non-renormalizable terms
of the form $S(\bar{\phi}\phi)^n/M_S^{2(n-1)}$, $n\geq 2$, are
also allowed \cite{jean} ($M_S\approx 5\times 10^{17}~{\rm GeV}$
is the string scale). The leading term of this type
$S(\bar{\phi}\phi)^2/M_S^2$, if its dimensionless coefficient is
of order unity, can be comparable to $S\bar{\phi}\phi$ (recall
that $\kappa\sim 10^{-3}$) and, thus, play a role in inflation
(see Sec.~\ref{sec:extensions}). All higher order terms of
this type with $n\geq 3$ give negligible contributions to the
inflationary potential (provided that $\vert\bar{\phi}\vert,~
\vert\phi\vert\ll M_S$ during inflation). The symmetry
${\rm U}(1)_{\rm R}$ guarantees \cite{lr} the linearity of $W$
in $S$ to all orders excluding terms such as $S^2$ which could
generate an inflaton mass $\stackrel{_{>}}{_{\sim }}H$ and ruin
inflation by violating the slow roll conditions.
\vspace{.25cm}
\item[{\bf iii.}] SUSY guarantees that the
radiative corrections do not ruin \cite{nonsinglet} inflation, but
rather provide \cite{dss} the necessary slope on the inflationary
path.
\vspace{.25cm}
\item[{\bf iv.}] Supergravity (SUGRA) corrections can be brought
under control leaving inflation intact. The scalar potential in
SUGRA is given by
\begin{equation}
V=\exp\left(\frac{K}{m_{\rm P}^2}\right)\left[\left(
K^{-1}\right)_i^{~j}F^i F_j-3\frac{\left|W\right|^2}
{m_{\rm P}^2}\right],
\label{eq:sugra}
\end{equation}
where $K$ is the K\"ahler potential, $m_{\rm P}=M_{\rm P}/
\sqrt{8\pi}\approx 2.44\times 10^{18}~{\rm GeV}$ is the reduced
Planck mass scale, $F^i=W^i+K^iW/m_{\rm P}^2$, and upper (lower)
indices denote derivation with respect to the scalar
field $\phi_i$ ($\phi^{j*}$). The K\"ahler potential can be
readily expanded as $K=\vert S\vert^2+\vert\bar{\phi}\vert^2+
\vert\phi\vert^2+\alpha\vert S\vert^4/m_{\rm P}^2+\cdots$, where
the quadratic terms constitute the minimal K\"ahler potential. The
term $\vert S\vert^2$, whose coefficient is normalized to unity,
could generate a ${\rm mass}^2\sim\kappa^2M^4/m_{\rm P}^2\sim H^2$
for $S$ on the inflationary path from the expansion of the
exponential prefactor in Eq.~(\ref{eq:sugra}). This would ruin
inflation. Fortunately, with this form of $W$ (including all the
higher order terms), this ${\rm mass}^2$ is cancelled in $V$
\cite{lyth,stewart}. The linearity of $W$ in $S$, guaranteed to
all orders by ${\rm U}(1)_{\rm R}$, is crucial for this
cancellation. The $\vert S\vert^4$ term
in $K$ also generates a ${\rm mass}^2$ for $S$ via the factor
$(\partial^2 K/\partial S\partial S^*)^{-1}=1-4\alpha
\vert S\vert^2/m_{\rm P}^2+\cdots$ in Eq.~(\ref{eq:sugra}),
which is however not cancelled (see e.g. Ref.~\cite{quasi}).
In order to avoid ruining inflation, one
has then to assume \cite{sugra,lss} that $|\alpha|
\stackrel{_{<}}{_{\sim }}10^{-3}$. All other higher
order terms in $K$ give suppressed contributions on the
inflationary path (since $\vert S\vert\ll m_{\rm P}$). So, we
see that a mild tuning of just one parameter is adequate for
controlling SUGRA corrections. (In other models, tuning of
infinitely many parameters is required.) Moreover, note that
with special forms of $K$ one can solve this problem even
without a mild tuning. An example is given in Ref.~\cite{costas},
where the dangerous ${\rm{mass}}^2$ term could be cancelled to all
orders in the presence of fields without superpotential but with
large VEVs generated via D-terms. Such a mechanism is necessary
in variants of hybrid inflation (see Ref.~\cite{nshift}) where
inflation takes place at large values of the inflaton field $S$
($\vert S\vert\sim m_{\rm P}$) since the higher order terms are,
in this case, unsuppressed and thus tuning of an infinite number
of parameters would be otherwise required. All the above methods
for controlling the SUGRA corrections also apply \cite{nshift} to
the extensions of the model that we will consider in
Sec.~\ref{sec:extensions}.
\end{list}
In summary, for all these reasons, we consider SUSY hybrid
inflation (with its extensions) as an extremely natural
inflationary scenario.

\section{Extensions of Supersymmetric Hybrid
Inflation} \label{sec:extensions}

Applying (SUSY) hybrid inflation to higher GUT gauge groups
predicting magnetic monopoles, we encounter the following
cosmological problem. Inflation is terminated abruptly as the
system reaches the critical point and is followed by the
waterfall regime during which the scalar fields $\bar\phi$,
$\phi$ develop their VEVs starting from zero and the spontaneous
breaking of the GUT gauge symmetry takes place. The fields
$\bar\phi$, $\phi$ can end up at any point of the vacuum manifold
with equal probability and, thus, magnetic monopoles are copiously
produced \cite{smooth} via the Kibble mechanism \cite{kibble}
leading to a disaster.

\par
One of the simplest GUTs predicting magnetic monopoles is the
Pati-Salam (PS) model \cite{ps} with gauge group $G_{\rm PS}=
{\rm SU}(4)_c\times {\rm SU}(2)_{\rm L}\times
{\rm SU}(2)_{\rm R}$. These monopoles carry \cite{magg} two
units of Dirac magnetic charge. We will present solutions
\cite{smooth,jean} of the monopole problem of hybrid inflation
within the SUSY PS model, although our mechanisms can be extended
to other gauge groups such as the trinification group
${\rm SU}(3)_c\times {\rm SU}(3)_{\rm L}\times
{\rm SU}(3)_{\rm R}$ (see e.g. Ref.~\cite{supuni}), which predicts
\cite{trinification} magnetic monopoles with triple Dirac charge.

\subsection{Shifted Hybrid Inflation}
\label{subsec:shifted}

One idea \cite{jean} for solving the monopole problem is to
include into the standard superpotential for hybrid inflation (in
Eq.~(\ref{eq:superpot})) the leading non-renormalizable term, which,
as explained, cannot be excluded. If its dimensionless coefficient
is of order unity, this term competes with the trilinear term of
the standard superpotential (with coefficient $\sim 10^{-3}$). A
totally new picture then emerges. There appears a non-trivial flat
direction along which $G_{\rm PS}$ is broken with the appropriate
Higgs fields acquiring constant values. This `shifted' flat
direction acquires a slope again from radiative corrections
\cite{dss} and can be used as inflationary path. The end of
inflation is again abrupt followed by a waterfall but no magnetic
monopoles are formed since $G_{\rm PS}$ is already broken during
inflation.

\par
The spontaneous breaking of the gauge group $G_{\rm PS}$ to
$G_{\rm S}$ is achieved via the VEVs of a conjugate pair of Higgs
superfields
\begin{eqnarray}
\bar{H}^c &=& (4,1,2)\equiv \left(\begin{array}{cccc} \bar{u}^c_H
& \bar{u}^c_H &
\bar{u}^c_H & \bar{\nu}_H^c\\
\bar{d}^c_H & \bar{d}^c_H & \bar{d}^c_H & \bar{e}^c_H
\end{array}\right),
\nonumber\\
H^c &=& (\bar{4},1,2)\equiv \left(\begin{array}{cccc}
u^c_H & u^c_H & u^c_H & \nu_H^c\\
d^c_H & d^c_H & d^c_H & e^c_H
\end{array}\right)
\label{eq:higgs}
\end{eqnarray}
in the $\bar{\nu}_H^c$, $\nu_H^c$ directions. The relevant part of
the superpotential, which includes the leading non-renormalizable
term, is
\begin{equation}
\delta W=\kappa S(-M^2+\bar{H}^c H^c)- \beta\frac{S(\bar{H}^c
H^c)^2}{M_S^2},
\label{eq:susyinfl}
\end{equation}
where $\beta$ is taken to be real and positive for simplicity.
D-flatness implies that $\bar{H}^{c} \,^{*}=e^{i\theta}H^c$. We
restrict ourselves to the direction with $\theta=0$
($\bar{H}^{c} \,^{*}=H^c$) which contains the shifted inflationary
trajectory (see below). The scalar potential derived from the
superpotential $\delta W$ in Eq.~(\ref{eq:susyinfl}) then takes
the form
\begin{equation}
V=\left[\kappa(\vert H^c\vert^2-M^2)- \beta\frac{\vert
H^c\vert^4}{M_S^2}\right]^2+ 2\kappa^2\vert S\vert^2 \vert
H^c\vert^2\left[1- \frac{2\beta}{\kappa M_S^2}\vert
H^c\vert^2\right]^2.
\label{eq:inflpot}
\end{equation}
Defining the dimensionless variables $w=\vert S \vert/M$, $y=\vert
H^c\vert/M$, we obtain
\begin{equation}
\tilde{V}=\frac{V}{\kappa^2M^4}=(y^2-1-\xi y^4)^2+ 2w^2y^2(1-2\xi
y^2)^2,
\label{eq:vtilde}
\end{equation}
where $\xi=\beta M^2/\kappa M_S^2$. This potential is a simple
extension of the standard potential for SUSY hybrid inflation
(which corresponds to $\xi=0$).

\par
For constant $w$ (or $|S|$), $\tilde V$ in Eq.~(\ref{eq:vtilde})
has extrema at
\begin{equation}
y_1=0,~y_2=\frac{1}{\sqrt{2\xi}},~y_{3\pm}=
\frac{1}{\sqrt{2\xi}}\sqrt{(1-6\xi w^2)\pm \sqrt{(1-6\xi
w^2)^2-4\xi(1-w^2)}}.
\label{eq:extrema}
\end{equation}
The first two extrema (at $y_1$, $y_2$) are $|S|$-independent and,
thus, correspond to classically flat directions, the trivial one at
$y_1=0$ with $\tilde{V}_1=1$ and the shifted one at
$y_2=1/\sqrt{2\xi}={\rm constant}$ with
$\tilde{V}_2=(1/4\xi-1)^2$, which we will use as inflationary
trajectory. The trivial trajectory is a valley of minima for
$w>1$, while the shifted one for $w>w_0=(1/8\xi-1/2)^{1/2}$, which
is its critical point. We take $\xi<1/4$, so that $w_0>0$ and the
shifted path is destabilized before $w$ reaches zero. The
extrema at $y_{3\pm}$, which are $|S|$-dependent and non-flat, do
not exist for all values of $w$ and $\xi$, since the expressions
under the square roots in Eq.~(\ref{eq:extrema}) are not always
non-negative. These two extrema, at $w=0$, become SUSY vacua. The
relevant SUSY vacuum (see below) corresponds to $y_{3-}(w=0)$ and,
thus, the common VEV $v_0$ of $\bar{H}^{c}$, $H^c$ is given by
\begin{equation}
\left(\frac{v_0}{M}\right)^2= \frac{1}{2\xi}(1-\sqrt{1-4\xi}).
\label{eq:v0}
\end{equation}

\par
We will now discuss the structure of $\tilde{V}$ and the
inflationary history for $1/6<\xi<1/4$. For fixed $w>1$, there
exist two local minima at $y_1=0$ and $y_2=1/\sqrt{2\xi}$, which
has lower potential energy density, and a local maximum at
$y_{3+}$ between the minima. As $w$ becomes smaller than unity,
the extremum at $y_1$ turns into a local maximum, while the
extremum at $y_{3+}$ disappears. The system then falls into the
shifted path in case it had started at $y_1=0$. As we further
decrease $w$ below $(2-\sqrt{36\xi-5})^{1/2}/3\sqrt{2\xi}$, a pair
of new extrema, a local minimum at $y_{3-}$ and a local maximum at
$y_{3+}$, are created between $y_1$ and $y_2$. As $w$ crosses
$(1/8\xi-1/2)^{1/2}$, the local maximum at $y_{3+}$ crosses $y_2$
becoming a local minimum. At the same time, the local minimum at
$y_2$ turns into a local maximum and inflation ends with the
system falling into the local minimum at $y_{3-}$ which, at $w=0$,
becomes the SUSY vacuum.

\par
We see that, no matter where the system starts from, it passes
from the shifted path, where the relevant part of inflation
takes place. So, $G_{\rm PS}$ is broken during inflation and no
magnetic monopoles are produced at the waterfall.

\par
After the termination of inflation, the system could fall into the
minimum at $y_{3+}$ instead of the one at $y_{3-}$. This, however,
does not happen since in the last e-folding or so the barrier
between the minima at $y_{3-}$ and $y_2$ is considerably reduced
and the decay of the `false vacuum' at $y_2$ to the minimum at
$y_{3-}$ is completed within a fraction of an e-folding before the
$y_{3+}$ minimum even appears.

\par
The only mass splitting within supermultiplets on the shifted
path appears \cite{jean} between one Majorana fermion in the
direction $(\bar{\nu}_H^c+\nu_H^c)/\sqrt{2}$ with $m^2=
4\kappa^2\vert S\vert^2$ and two real scalar fields
$\mathrm{Re}(\delta\bar{\nu}^c_H+\delta \nu^c_H)$ and
$\mathrm{Im}(\delta \bar{\nu}^c_H+\delta\nu^c_H)$ with
$m_{\pm}^2=4\kappa^2\vert S\vert^2\mp 2 \kappa^2m^2$. Here,
$m=M(1/4\xi-1)^{1/2}$ and $\delta\bar{\nu}^c_H=\bar{\nu}^c_H-v$,
$\delta\nu^c_H=\nu^c_H-v$, where $v=(\kappa M_S^2/2\beta)^{1/2}$
is the value of $\bar{H}^c$, $H^c$ on the shifted inflationary
path.

\par
The radiative corrections on the shifted inflationary trajectory
can be readily constructed and $(\delta T/T)_Q$ and $\kappa$ can
be evaluated. We find the same formulas as in Eqs.~(\ref{eq:qa})
and (\ref{eq:kappa}) with $\mathcal{N}=2$ and $\mathcal{N}=4$
respectively and $M$ generally replaced by $m$. The COBE results
\cite{cobe} can be reproduced, for instance, with $\kappa
\approx 4\times 10^{-3}$, corresponding to $\xi=1/5$, $v_0\approx
1.7\times 10^{16}~{\rm GeV}$ ($N_Q\approx 55$, $\beta=1$). The
scales $M\approx 1.45\times 10^{16}~{\rm GeV}$, $m\approx 7.23
\times 10^{15}~{\rm GeV}$, the inflaton mass $m_{\mathrm{infl}}
\approx 4.1\times 10^{13} ~{\rm GeV}$, and the inflationary
scale, which characterizes the inflationary vacuum energy
density, $v_{\rm infl}=\kappa^{1/2}m \approx 4.57\times 10^{14}
~{\rm GeV}$. The spectral index $n_s\approx 0.985$ \cite{senoguz}.
It is interesting to note that this scenario can also be realized
\cite{nshift} with only renormalizable superpotential couplings.

\subsection{Smooth Hybrid Inflation}
\label{subsec:smooth}

\par
An alternative solution to the magnetic monopole problem of hybrid
inflation has been proposed in Ref.~\cite{smooth}. We will present
it here within the SUSY PS model of Sec.~\ref{subsec:shifted},
although it can be applied to other semi-simple gauge groups too.
The idea is to impose an extra $Z_2$ symmetry under which
$H^c\rightarrow -H^c$. The whole structure of the model remains
unchanged except that now only even powers of the combination
$\bar{H}^cH^c$ are allowed in the superpotential terms.

\par
The inflationary superpotential in Eq.~(\ref{eq:susyinfl}) becomes
\begin{equation}
\delta W=S\left(-\mu^2+\frac{(\bar{H}^cH^c)^2} {M_S^2}\right),
\label{eq:smoothsuper}
\end{equation}
where we absorbed the dimensionless parameters $\kappa$, $\beta$
in $\mu$, $M_S$. The resulting scalar potential $V$ is then given
by
\begin{equation}
\tilde{V}=\frac{V}{\mu^4}=(1-\tilde\chi^4)^2+
16\tilde\sigma^2\tilde\chi^6,
\label{eq:smoothpot}
\end{equation}
where we used the dimensionless fields $\tilde\chi=\chi/2(\mu
M_S)^{1/2}$, $\tilde\sigma =\sigma/2(\mu M_S)^{1/2}$ with $\chi$,
$\sigma$ being normalized real scalar fields defined by
$\bar{\nu}_H^c=\nu_H^c=\chi/2$, $S=\sigma/\sqrt{2}$ after rotating
$\bar{\nu}_H^c$, $\nu_H^c$, $S$ to the real axis.

\par
The emerging picture is completely different. The flat direction
at $\tilde\chi=0$ is now a local maximum with respect to
$\tilde\chi$ for all values of $\tilde\sigma$ and two new
symmetric valleys of minima appear \cite{smooth,shi} at
\begin{equation}
\tilde\chi=\pm\sqrt{6}\tilde\sigma\left[\left(1+
\frac{1}{36\tilde\sigma^4}\right)^{\frac{1}{2}}-1
\right]^{\frac{1}{2}}.
\label{eq:smoothvalley}
\end{equation}
They contain the SUSY vacua lying at $\tilde\chi= \pm 1$,
$\tilde\sigma=0$ and possess a slope already at the classical
level. So, in this case, there is no need of radiative corrections
for driving the inflaton. The potential on these paths is
\cite{smooth,shi}
\begin{equation}
\tilde{V}=48\tilde\sigma^4 \left[72\tilde\sigma^4\left(1+
\frac{1}{36\tilde\sigma^4}\right)
\left(\left(1+\frac{1}{36\tilde\sigma^4}
\right)^{\frac{1}{2}}-1\right)-1\right].
\label{eq:smoothV}
\end{equation}
The system follows a particular inflationary path and ends up at a
particular point of the vacuum manifold leading to no production
of monopoles.

\par
The end of inflation is not abrupt since the inflationary path is
stable with respect to $\tilde\chi$ for all $\tilde\sigma$'s. It
is determined by using the $\epsilon$ and $\eta$ criteria.
Moreover, as it has been shown \cite{smoothini}, the initial
values (at $t_{\rm P}$) of the fields which can lead to adequate
inflation in this model are less restricted than in the standard
SUSY hybrid inflationary scenario.

\par
This model allows us to take the VEV $v_0=(\mu M_S)^{1/2}$ of
$\bar{H}^c$, $H^c$ equal to the SUSY GUT VEV. COBE \cite{cobe}
then yields $M_S\approx 7.87\times 10^{17} ~{\rm GeV}$ and
$\mu\approx 1.04\times 10^{15} ~{\rm GeV}$ for $N_Q\approx 57$.
Inflation ends at $\sigma=\sigma_0\approx 1.08\times 10^{17}~{\rm
GeV}$, while our present horizon crosses outside the inflationary
horizon at $\sigma= \sigma_Q\approx 2.72\times 10^{17}~{\rm GeV}$.
Finally, $m_{\rm infl}=2\sqrt{2}\mu^2/v_0 \approx 1.07\times
10^{14}~{\rm GeV}$ and the spectral index $n_s\approx 0.97$.

\section{Conclusions}
\label{sec:conclusions}

\par
We summarized the shortcomings of the SBB cosmological model and
their resolution by inflation, which suggests that the universe
underwent a period of exponential expansion. This may have
happened during the GUT phase
transition at which the relevant Higgs field was displaced from
the vacuum. This field (inflaton) could then, for some time, roll
slowly towards the vacuum providing an almost constant vacuum
energy density. Inflation generates the density perturbations
needed for the large scale structure of the universe and the
temperature fluctuations of the CMBR. After the end of inflation,
the inflaton performs damped oscillations about the vacuum,
decays, and reheats the universe.

\par
The early inflationary models required tiny parameters. This
problem was solved by hybrid inflation which uses two real scalar
fields. One of them provides the vacuum energy density for
inflation while the other one is the slowly rolling field. Hybrid
inflation arises naturally in many SUSY GUTs, but leads to a
disastrous overproduction of magnetic monopoles. We constructed
two extensions of SUSY hybrid inflation which do not suffer from
this problem.

\section*{Acknowledgements}
This work was supported by the European Union under the contract
MRTN-CT-2004-503369.

\def\plb#1#2#3{#3~{\it Phys. Lett.} B
{\bf #1}~#2}
\def\apjl#1#2#3{#3~{\it Astrophys. J. Lett.}
{\bf #1}~#2}
\def\apj#1#2#3{#3~{\it Astrophys. J.}
{\bf #1}~#2}
\def\jetpl#1#2#3{#3~{\it JETP Lett.}
{\bf #1}~#2}
\def\jetpsp#1#2#3{#3~{\it JETP (Sov. Phys.)}
{\bf #1}~#2}
\def\spss#1#2#3{#3~{\it Sov. Phys. -Solid State}
{\bf #1}~#2}
\def\jpa#1#2#3{#3~{\it J. Phys.} A
{\bf #1}~#2}
\def\pr#1#2#3{#3~{\it Phys. Reports}
{\bf #1}~#2}
\def\mnras#1#2#3{#3~{\it Mon. Not. Roy. Astr. Soc.}
{\bf #1}~#2}
\def\n#1#2#3{#3~{\it Nature}
{\bf #1}~#2}
\def\cmp#1#2#3{#3~{\it Commun. Math. Phys.}
{\bf #1}~#2}
\def\prsla#1#2#3{#3~{\it Proc. Roy. Soc. London} A
{\bf #1}~#2}
\def\ptp#1#2#3{#3~{\it Prog. Theor. Phys.}
{\bf #1}~#2}
\def\prd#1#2#3{#3~{\it Phys. Rev.} D
{\bf #1}~#2}
\def\prl#1#2#3{#3~{\it Phys. Rev. Lett.}
{\bf #1}~#2}
\def\npb#1#2#3{#3~{\it Nucl. Phys.} B
{\bf #1}~#2}
\def\jhep#1#2#3{#3~{\it J. High Energy Phys.}
{\bf #1}~#2}
\def\anj#1#2#3{#3~{\it Astron. J.}
{\bf #1}~#2}
\def\baas#1#2#3{#3~{\it Bull. Am. Astron. Soc.}
{\bf #1}~#2}
\def\grg#1#2#3{#3~{\it Gen. Rel. Grav.}
{\bf #1}~#2}
\def\stmp#1#2#3{#3~{\it Springer Trac. Mod. Phys.}
{\bf #1}~#2}
\def\ijmpa#1#2#3{#3~{\it Int. J. Mod. Phys.} A
{\bf #1}~#2}
\def\lnp#1#2#3{#3~{\it Lect. Notes Phys.}
{\bf #1}~#2}
\def\apjs#1#2#3{#3~{\it Astrophys. J. Suppl.}
{\bf #1}~#2}
\def\jcap#1#2#3{#3~{\it J. Cosmol. Astropart. Phys.}
{\bf #1}~#2}
\def\ijmpe#1#2#3{#3~{\it Int. J. Mod. Phys.} E
{\bf #1}~#2}
\def\aap#1#2#3{#3~{\it Astron. Astrophys.}
{\bf #1}~#2}
\def\epjc#1#2#3{#3~{\it Eur. Phys. J.} C
{\bf #1}~#2}

\end{document}